\documentclass[twocolumn,showpacs,preprintnumbers]{revtex4}
\usepackage{psfig}
\begin{document}

\preprint{FZJ-IKP-TH-2008-8, HISKP-TH-08/06}

\title{
Near threshold {\boldmath $p{\bar p}$} enhancement in 
the {\boldmath$J/\psi \to \omega p{\bar p}$} decay}

\author{J. Haidenbauer$^1$, Ulf-G. Mei{\ss}ner$^{2,1}$, A. Sibirtsev$^{1,2}$}

\affiliation{
$^1$Institut f\"ur Kernphysik (Theorie), Forschungszentrum J\"ulich,
D-52425 J\"ulich, Germany \\
$^2$Helmholtz-Institut f\"ur Strahlen- und Kernphysik (Theorie), 
Universit\"at Bonn, Nu\ss allee 14-16, D-53115 Bonn, Germany 
}

\begin{abstract}
The near-threshold behavior of the $p{\bar p}$ invariant mass spectrum 
from the $J/\psi{\to}\omega p{\bar p}$ decay reported recently by the 
BES Collaboration is analyzed. 
Contrary to the statement made by the BES Collaboration itself
our study demonstrates that there is indeed a noticeable enhancement 
in the $p{\bar p}$ invariant mass spectrum near threshold. 
Moreover, this enhancement is nicely reproduced by the 
final state interaction in the relevant ($^{11}S_0$)
$p{\bar p}$ partial wave as given by the J\"ulich nucleon--antinucleon
model. Therefore, and again
contrary to the statement by the BES Collaboration,
their new data on $J/\psi {\to}\omega p{\bar p}$ decay
in fact strongly support the FSI interpretation of the $p{\bar p}$ 
enhancement, seen also in other decay reactions.
\end{abstract}

\maketitle

The study of the decays of mesons like the $J/\psi$, $\psi (2S)$,
$B$, and $\Upsilon$ as pursued by the BES, Belle, BABAR and
CLEO Collaborations is a rather powerful tool for 
examining systematically the spectrum of light as well
as heavier hadrons. Specifically, exclusive measurements of 
decays into three-meson or meson-baryon-antibaryon channels
play a very important role and have already led to the
identification of several new structures. 
 
Among the various three-particle channels explored those involving 
the proton-antiproton ($p\bar p$) system in the final state
have caused considerable attention in the community. 
The excitement was initiated by the observation of a significant 
near-threshold enhancement in the $p{\bar p}$ invariant mass 
spectrum for the reaction $J/\psi{\to}\gamma p{\bar p}$ 
in a high-statistics and high-mass-resolution experiment
by the BES Collaboration~\cite{Bai}.
Indeed a first indication for a near-threshold enhancement in the 
$p{\bar p}$ invariant mass spectrum from the $B^+{\to}K^+p{\bar p}$ 
and ${\bar B^0}{\to}D^0p{\bar p}$ decays were reported 
by the Belle Collaboration~\cite{Abe1,Abe2} but with much
lower statistics and mass resolution. 
More recently the Belle Collaboration~\cite{Wang,Wei}
found also a near-threshold $p{\bar p}$ enhancement in 
the decays $B^+{\to}\pi^+p{\bar p}$,
$B^0{\to}K^0p{\bar p}$ and $B^+{\to}K^{\ast
+}p{\bar p}$, while the CLEO Collaboration detected such
an enhancement in (the unsubtracted) data for 
$\Upsilon (1S) \to \gamma p{\bar p}$ \cite{Cleo} 
and the BES Collaboration in 
$\psi (2S) \to \gamma p{\bar p}$ \cite{Ablikim}.
Finally, the BABAR Collaboration presented measurements of 
the $B^+{\to}K^+p{\bar p}$, $B^0{\to}\bar D^0p{\bar p}$ and
$B^0{\to}\bar D^{*0}p{\bar p}$ decays~\cite{Aubert,Aubert1} 
confirming the presence of a near-threshold enhancement in 
the $p{\bar p}$ invariant mass.
 
The high-statistics data by the BES Collaboration triggered several 
theoretical speculations where the observed enhancement in the 
invariant $p{\bar p}$ mass spectrum was interpreted as evidence for a 
$p{\bar p}$ bound state or baryonium~\cite{Datta,Ding1,Ding2,Suzuki}, 
or for exotic glueball states~\cite{Chua,Rosner}. 
Alternatively, we \cite{Sibirtsev1,Sibirtsev2} but also others 
\cite{Kerbikov,Bugg,Zou,Loiseau,Entem,Laporta} demonstrated that the 
near-threshold enhancement in the $p{\bar p}$ invariant mass 
spectrum from $J/\psi{\to}\gamma p{\bar p}$ and other decays leading
to a final $p{\bar p}$ system 
could be simply due to the final state interaction (FSI) between the 
outgoing proton and antiproton. Specifically, our calculation based 
on the realistic J\"ulich nucleon--antinucleon
($N{\bar N}$) model~\cite{Hippchen,Mull},
the one by Loiseau and Wycech \cite{Loiseau}, utilizing the Paris
$N{\bar N}$ model, and those of Entem and Fern\'andez \cite{Entem},
using a $N{\bar N}$ interaction derived from a constituent quark model, 
explicitly confirmed the significance of FSI 
effects estimated in the initial studies~\cite{Kerbikov,Bugg,Zou}
within the effective range approximation. Interestingly, the
same FSI mechanism explains the near threshold enhancement of the data
on $e^+ e^- \leftrightarrow \bar NN $ from the PS170 collaboration, from 
the FENICE collaboration and from BABAR utilizing radiative return, 
see~\cite{Sibirtsev3,Dmitriev}.

Very recently the BES Collaboration presented
a high-statistics measurement of the $J/\psi {\to}\omega p{\bar p}$ 
decay~\cite{BES07} where, according to their own words, no
obvious near-threshold $p{\bar p}$ mass enhancement is observed. 
This supposed lack of any enhancement is then seen as a hint that
the FSI interpretation of the $p{\bar p}$ enhancement in 
$J/\psi {\to}\gamma p{\bar p}$ is disfavoured \cite{BES07}. 

In the present paper we want to take a closer look at those
$J/\psi {\to}\omega p{\bar p}$ data by the BES Collaboration.
As we already argued in our first work on the $p{\bar p}$ enhancement
\cite{Sibirtsev1}, this specific decay channel is rather interesting
for clarifying the role of the $p{\bar p}$ FSI effects, because here
the conservation laws for parity, charge-conjugation and total 
angular momentum severely restrict the 
partial waves in the $p{\bar p}$ system. In particular, near
threshold the $p{\bar p}$ system can only be in the 
$^{11}S_0$ state. We use here the standard nomenclature $^{(2I+1)(2S+1)}L_J$ 
where $I$ and $S$ are the total isospin and spin, respectively. 
In contrary, for the extensively discussed $J/\psi{\to}\gamma p{\bar p}$ 
decay any combination of the $I=0$ and $I=1$ amplitudes 
is allowed because isospin is not conserved in electromagnetic processes. 

\begin{figure}[t]
\vspace*{-5mm}
\centerline{\hspace*{3mm}\psfig{file=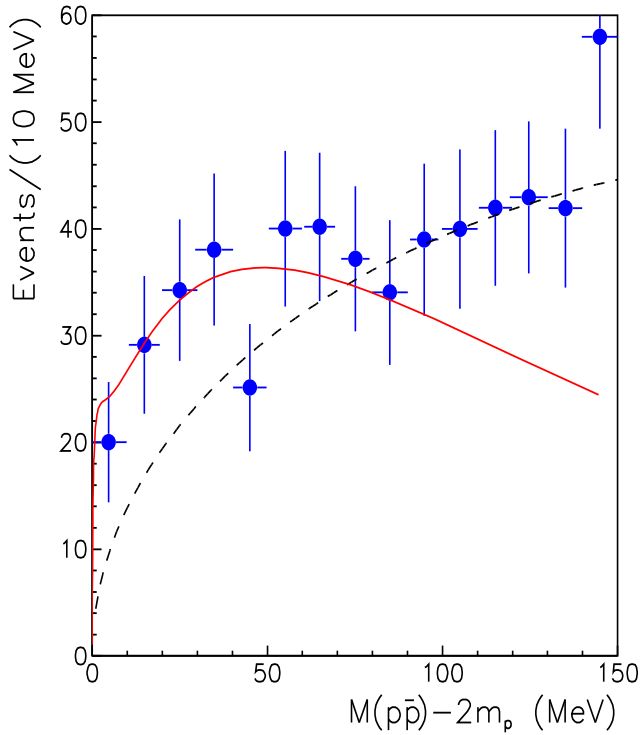,width=9.7cm,height=11.cm}}
\vspace*{-5mm}
\caption{
The $p\bar p$ mass spectrum from the decay $J/\psi{\to}\omega p{\bar p}$.
The circles show experimental results of the BES Collaboration 
\cite{BES07}, while the dashed line is the spectrum obtained from Eq. (\ref{trans}) 
by assuming a constant reaction amplitude $A$ which was normalized to the
data at $M(p\bar p)-2m_p\approx$ 110 MeV. 
The solid line is a calculation using the scattering amplitude squared 
($|T|^2$) predicted by the $N\bar N$ model A(OBE) \cite{Hippchen}
for the $^{11}S_0$ partial wave, normalized to the data, cf.
Eq.~(\ref{fsi}). 
}
\label{besnew2}
\end{figure}

\begin{figure}[t]
\vspace*{-5mm}
\centerline{\hspace*{3mm}\psfig{file=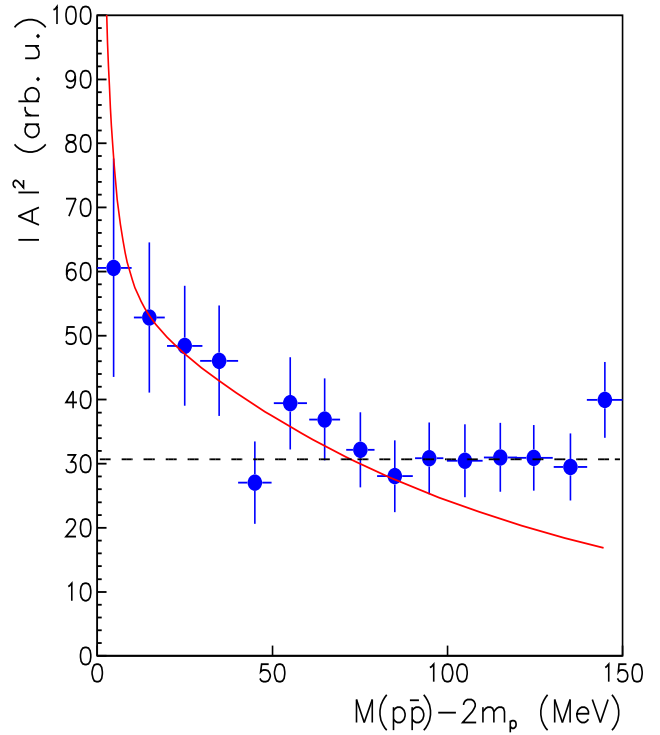,width=9.7cm,height=11.cm}}
\vspace*{-5mm}
\caption{
Invariant $J/\psi{\to}\omega p{\bar p}$ amplitude $|A|^2$ as a
function of the $p{\bar p}$ mass. The circles symbolize the experimental
values of $|A|^2$ extracted from the BES data \cite{BES07} 
via Eq.~(\ref{trans}). The solid curve is the appropriately normalized 
scattering amplitude squared ($|T|^2$) predicted by the $N\bar N$ 
model A(OBE) \cite{Hippchen} for the $^{11}S_0$ partial wave. 
The dashed curve represents the constant reaction amplitude used
for generating the dashed curve in Fig.~\ref{besnew2}.
}
\label{besnew1}
\end{figure}

Like in our earlier papers \cite{Sibirtsev1,Sibirtsev2,Sibirtsev3}, 
besides the directly measured $p\bar p$ invariant mass spectrum, 
we utilize also the total spin-averaged (dimensionless) 
$J/\psi {\to}\omega p{\bar p}$ reaction amplitude $A$ 
because that allows us to get rid of trivial kinematical factors. 
The $J/\psi {\to}\omega p{\bar p}$ decay rate is given in terms of $A$ 
by~\cite{Byckling}
\begin{eqnarray}
d\Gamma = \frac{|A|^2}{2^9 \pi^5 m_{J/\psi}^2}\,
\lambda^{1/2}(m_{J/\psi}^2,M^2,m_{\omega}^2) \nonumber \\
\times\lambda^{1/2}(M^2,m_p^2,m_p^2)\, dM d\Omega_p\,  d\Omega_\omega,
\label{spectr}
\end{eqnarray}
where the Kallen function $\lambda$ is defined by
$\lambda (x,y,z)={((x-y-z)^2-4yz})/{4x}\,$,
$M \equiv M(p\bar p)$  is the invariant mass of the $p{\bar p}$ 
system, $\Omega_p$ is the proton angle in that system, 
while $\Omega_\omega$ is the $\omega$ angle in
the $J/\psi$ rest frame. After averaging over the spin states and
integrating over the angles, the differential decay rate is
\begin{eqnarray}
\frac{d\Gamma}{dM}=\frac{\lambda^{1/2}(m_{J/\psi}^2,M^2,m_{\omega}^2)
\sqrt{M^2-4m_p^2}}{2^6 \pi^3 m_{J/\psi}^2}\,\, |A|^2 \ .
\label{trans}
\end{eqnarray}
We use Eq.~(\ref{trans}) for extracting $|A|^2$ from the data of the
BES Collaboration. The original data \cite{BES07} are reproduced in 
Fig.~\ref{besnew2} while the extracted values for $|A|^2$ are shown 
in Fig.~\ref{besnew1}. 

We 
assume again the validity of the Watson-Migdal approach for the
treatment of the FSI effect. It suggests that the reaction amplitude
for a production and/or decay reaction that is of short-ranged
nature can be factorized in terms of an elementary (basically
constant) production amplitude and the $p\bar p$ scattering 
amplitude $T$ of the particles in the final state so that
\begin{eqnarray}
A (M(p \bar p)) \approx N \cdot T(M(p \bar p)), 
\label{fsi}
\end{eqnarray}
(cf. Ref. \cite{Sibirtsev1} for further details). 
Thus, we compare the extracted amplitude $|A|^2$ with the 
suitably normalized scattering amplitudes $|T|^2$ that 
result from the J\"ulich $N\bar N$ model \cite{Hippchen} 
for the $^{11}S_0$ partial wave. Interestingly, that 
scattering amplitude reproduces the dependence of the 
experimental $|A|^2$ on the invariant mass almost perfectly 
in the near-threshold region, cf. the solid curve in 
Fig.~\ref{besnew1}.
Therefore, we feel that we even need to stress that this 
result is actually a prediction of the model and not a fit. 
The dashed line represents a constant reaction 
amplitude and corresponds to the pure phase-space behavior. 
Obviously the BES data show a clear enhancement 
as compared to the phase-space behavior in the near-threshold
region, contrary to what is concluded by the authors from 
these data in their own publication \cite{BES07}. 
This can be also seen from Fig.~\ref{besnew2}, where
the measured $p{\bar p}$ mass spectrum is shown directly. 
The normalization of the phase space is done in the region
$M(p{\bar p})-2m_p\approx$ 100-140 MeV, where the data indeed follow 
the phase-space distribution. In principle, one could have also
normalized the dashed curve to the lowest data points. Then the 
first four data points would still be roughly in line 
with a phase-space behavior, at least within the error bars,
but one would end up with a gross overestimation of the
data at higher invariant masses and, consequently, be in a 
situation that one sees and has to explain a suppression 
in the experimental data in that invariant-mass region. 

Note that the disagreement of our model results with the experiment 
for invariant masses beyond $M(p{\bar p})-2m_p\approx$ 100 MeV 
is not a reason of concern and, in particular, does not 
discredit the interpretation of the data in terms of FSI effects.
At those energies 
we expect that contributions from higher partial waves,
not considered here, should start to play a more prominent role. 
 
In summary, we have analyzed the near-threshold data on the
$p{\bar p}$ invariant mass spectrum from the 
$J/\psi {\to}\omega p{\bar p}$ decay reported recently by the 
BES Collaboration.
Contrary to the statement made by the BES Collaboration in
\cite{BES07} our study demonstrates that 
not only in $J/\psi {\to}\gamma p{\bar p}$ but 
also in this reaction there is indeed a noticeable enhancement 
in the $p{\bar p}$ invariant mass spectrum near threshold. 
Moreover, this enhancement is nicely reproduced by the 
final state interaction in the relevant ($^{11}S_0$)
$p{\bar p}$ partial wave as given by the J\"ulich $N\bar N$
model \cite{Hippchen}. Accordingly, the present result is 
completely in line with our previous 
investigations of the $p{\bar p}$ invariant mass spectrum from the 
$J/\psi{\to}\gamma p{\bar p}$ decay \cite{Sibirtsev1} measured by the
BES Collaboration and the $B^+{\to}K^+ p{\bar p}$ decay \cite{Sibirtsev2}
measured by the BABAR Collaboration. In particular, and again
contrary to the statement by the BES Collaboration \cite{BES07}, 
their new data on $J/\psi {\to}\omega p{\bar p}$ decay, in fact,
strongly support the FSI interpretation of the $p{\bar p}$ enhancement 
seen in other decay reactions.
It goes without saying that, the FSI effects for the various decay 
reactions should not be expected to be {\it quantitatively the same} 
because due to the different quantum numbers and 
conservation laws as well as different reaction mechanisms,
the final $p{\bar p}$ system can and must be in different 
partial waves.

\acknowledgments{
This work was partially supported by the 
DFG (SFB/TR 16, ``Subnuclear Structure of Matter''), by
the EU Integrated Infrastructure Initiative Had{\-}ron Physics Project 
(contract no. RII3-CT-2004-506078) and by the Helmholtz Association
through funds provided to the virtual institute ``Spin and strong
QCD'' (VH-VI-231). A.S. acknowledges support by
the JLab grant SURA-06-C0452 and the COSY FFE grant No. 41760632
(COSY-085).
}


\vfill

\begin{thebibliography}{99}
\bibitem{Bai}
        J.Z. Bai {\it et al.}, Phys. Rev. Lett. {\bf 91}, 022001 (2003)
        [arXiv:hep-ex/0303006].
\bibitem{Abe1}
        K. Abe {\it et al.}, Phys. Rev. Lett. {\bf 88}, 181803 (2002)
        [arXiv:hep-ex/0202017].
\bibitem{Abe2}
        K. Abe {\it et al.}, Phys. Rev. Lett. {\bf 89}, 151802 (2002)
        [arXiv:hep-ex/0205083].
\bibitem{Wang}
        M.Z. Wang {\it et al.}, Phys. Rev. Lett. {\bf 92}, 131801 (2004);
        M.Z. Wang {\it et al.}, Phys. Lett. B {\bf 617}, 141 (2005). 
\bibitem{Wei}
  J.~T.~Wei {\it et al.},
  Phys.\ Lett.\  B {\bf 659}, 80 (2008)
  [arXiv:0706.4167 [hep-ex]].
\bibitem{Cleo}
        S.B. Athar {\it et al.}, Phys. Rev. D {\bf 73}, 032001 (2006).
\bibitem{Ablikim}
  M.~Ablikim {\it et al.}, 
  Phys.\ Rev.\ Lett.\  {\bf 99}, 011802 (2007)
  [arXiv:hep-ex/0612016].

\bibitem{Aubert}
        B. Aubert {\it et al.}, Phys. Rev. D {\bf 72}, 051101 (R) (2005).
\bibitem{Aubert1}
  B. Aubert {\it et al.},
  Phys.\ Rev.\  D {\bf 74}, 051101 (2006)
  [arXiv:hep-ex/0607039].
\bibitem{Datta}
        A. Datta and P.J. O'Donnell, Phys. Lett. B {\bf 567}, 273 (2003).
\bibitem{Ding1}
  G.~J.~Ding and M.~L.~Yan,
  Phys.\ Rev.\  C {\bf 72}, 015208 (2005)
  [arXiv:hep-ph/0502127].
\bibitem{Ding2}
  G.~J.~Ding, J.~l.~Ping and M.~L.~Yan,
  Phys.\ Rev.\  D {\bf 74}, 014029 (2006)
  [arXiv:hep-ph/0510013].
\bibitem{Suzuki}
  M.~Suzuki,
  J.\ Phys.\ G {\bf 34}, 283 (2007)
  [arXiv:hep-ph/0609133].
\bibitem{Chua}
        C. K. Chua, W. S. Hou and S. Y. Tsai, Phys. Lett. B {\bf 544},
        139 (2002).
\bibitem{Rosner}
        J.L. Rosner, Phys. Rev. D {\bf 68}, 014004 (2003).
\bibitem{Sibirtsev1}
        A. Sibirtsev, J. Haidenbauer, S. Krewald, U.-G. Mei{\ss}ner
        and A.W. Thomas, Phys. Rev. D {\bf 71}, 054010 (2005)
        [arXiv:hep-ph/0411386].
\bibitem{Sibirtsev2}
  J.~Haidenbauer, U.-G.~Mei{\ss}ner and A.~Sibirtsev,
  Phys.\ Rev.\  D {\bf 74}, 017501 (2006)
  [arXiv:hep-ph/0605127].
\bibitem{Kerbikov}
        B. Kerbikov, A. Stavinsky, and V. Fedotov, Phys. Rev. C
        {\bf 69}, 055205 (2004) [arXiv:hep-ph/0402054].
\bibitem{Bugg}
        D.V. Bugg, Phys. Lett. B {\bf 598}, 8 (2004)
        [arXiv:hep-ph/0406293].
\bibitem{Zou}
        B.S. Zou and  H.C. Chiang, Phys. Rev. D {\bf 69},
        034004 (2004) [arXiv:hep-ph/0309273].
\bibitem{Loiseau}
        B. Loiseau and S. Wycech, Phys. Rev. C {\bf 72},
        011001 (2005) [arXiv:hep-ph/0501112].
\bibitem{Entem}
  D.~R.~Entem and F.~Fern{\'a}ndez,
  Phys.\ Rev.\  D {\bf 75}, 014004 (2007).
\bibitem{Laporta}
  V.~Laporta,
  Int.\ J.\ Mod.\ Phys.\  A {\bf 22}, 5401 (2007)
  [arXiv:0707.2751 [hep-ph]].

\bibitem{Hippchen}
        T. Hippchen, J. Haidenbauer, K. Holinde, V. Mull, 
        Phys. Rev. C {\bf 44}, 1323 (1991); 
        V. Mull, J. Haidenbauer, T. Hippchen, K. Holinde, 
        Phys. Rev. C {\bf 44}, 1337 (1991).
\bibitem{Mull}
        V. Mull, K. Holinde,
        Phys. Rev. C {\bf 51}, 2360 (1995).

\bibitem{Sibirtsev3}
  J.~Haidenbauer, H.~W.~Hammer, U.-G.~Mei{\ss}ner and A.~Sibirtsev,
  Phys.\ Lett.\  B {\bf 643}, 29 (2006)
  [arXiv:hep-ph/0606064].
\bibitem{Dmitriev}
  V.~F.~Dmitriev and A.~I.~Milstein,
  Phys.\ Lett.\  B {\bf 658}, 13 (2007).

\bibitem{BES07}
  M.~Ablikim {\it et al.}, 
  Eur.\ Phys.\ J.\  C {\bf 53}, 15 (2008)
  [arXiv:0710.5369 [hep-ex]].

\bibitem{Byckling}
        E. Byckling and K. Kajantie, Particle Kinematics, John
        Willey and Sons (1973).
\end{thebibliography}
\end{document}